# Three-dimensional Simulation of Quantitative Ultrasound in Cancellous Bone using the Echographic Response of a Metallic Pin

Yoshiki Nagatani[1]
    Department of Electronics, Kobe City College of Technology, 8-3 Gakuen-higashi-machi, Nishiku, 651-2194, Kobe, Japan / nagatani@ultrasonics.jp / +81-78-795-3247

Séraphin Guipieri, Vu-Hieu Nguyen
    Université Paris-Est, Laboratoire de Modélisation et Simulation Multi Echelle, UMR CNRS 8208, Créteil F-94010, France

Christine Chappard
    CNRS, B2OA UMR CNRS 7052, Paris, France

Didier Geiger, Salah Naili
    Université Paris-Est, Laboratoire de Modélisation et Simulation Multi Echelle, UMR CNRS 8208, Créteil F-94010, France

Guillaume Haïat[2]
    CNRS, Laboratoire de Modélisation et Simulation Multi Echelle, UMR CNRS 8208, Créteil F-94010, France

**Corresponding author**: Yoshiki Nagatani

**Keywords**: FDTD Simulation; Cancellous Bone; Quantitative ultrasound (QUS); 3D Model; Discopathy

1. Also at: Université Paris-Est, Laboratoire de Modélisation et Simulation Multi Echelle, UMR CNRS 8208, Créteil F-94010, France
2. Also at : École de technologie supérieure, 1100 Notre-Dame Street West, Montreal, QC H3C 1K3, Canada; Research Center, Hôpital du Sacré-Cœur de Montréal, 5400, Gouin Boul. West, Montreal, QC H4J 1C5, Canada.

**Acknowledgement**

The authors appreciate Clara Ceron and Hung-Son Phan, technicians of MSME laboratory, providing perfect computing environment. This study was supported in part by KAKENHI (Grant Number 25871038 and 16K01431) from the Japan Society for the Promotion of Science (JSPS) and the PRTS program (project OsseoWave n°ANR-13-PRTS-0015-02) of French National Research Agency (ANR). This paper has received funding from the European Research Council (ERC) under the European Union's Horizon 2020 research and innovation program (grant agreement No 682001, project ERC Consolidator Grant 2015 BoneImplant).

ABSTRACT

Degenerative discopathy is a common pathology which may require spine surgery. A metallic cylindrical pin is inserted into the vertebral body to maintain soft tissues and may be used as a reflector of ultrasonic wave to estimate bone density. The first aim of this paper is to validate a 3D model to simulate the ultrasonic propagation in a trabecular bone sample in which a metallic pin has been inserted. We also aim at determining the effect of changes of bone volume fraction (BV/TV) and of positioning errors on the quantitative ultrasound (QUS) parameters in this specific configuration. The approach consists in coupling finite difference time domain simulation with X-ray microcomputed tomography. The correlation coefficient between experimental and simulated speed of sound (SOS) (respectively broadband ultrasonic attenuation (BUA)) was equal to 0.90 (respectively 0.55). The results show a significant correlation of SOS with BV/TV ($R$ = 0.82), while BUA values exhibit a non-linear behavior *versus* BV/TV. The orientation of the pin should be controlled with an accuracy of around 1° in order to obtain accurate results. The results indicate that using the ultrasonic wave reflected by a pin has a potential to estimate the bone density. SOS is more reliable than BUA due its lower sensitivity to the tilt angle.

I. INTRODUCTION

Degenerative discopathy (DD) is one of the most common etiology of spinal degeneration and may require spine surgery [1]. Spinal fusion is one of the possible options to treat intervertebral disc (ID) degeneration. In spinal fusion surgery, various implantable devices are used to connect two adjacent vertebrae such as cages, plates, screws and a variety of other fusion materials. In some cases, external fixations (such as pedicle screws) may be used in order to reinforce the biomechanical stability of the vertebral structure [2]. However, the choice of the method used for pedicle screw insertion (use of cement, length and diameter of the screws) remains empirical and depends on bone biomechanical properties [3].

Another technique for the treatment of ID degeneration is total disc replacement (TDR), which has the advantage of preserving spinal motion [4] and maintaining a mobility between the two adjacent vertebrae [5]. However, the long-term stability of implants used in TDR remains difficult to obtain due to problems of bony fixation and anchorage to the host spine, which depends on the vertebral bone properties. Moreover, a risk of impaction fracture of the vertebrae still remains, depending on bone properties.

For all the aforementioned reasons, a reliable method to assess bone biomechanical properties would be of interest in spine surgery because the number of osteoporotic patients increases with age.

Dual-energy X-ray Absorptiometry (DEXA) is currently the only method available in clinical practice to estimate vertebral bone quantity and is sometimes used before spine surgery. It remains difficult to use DEXA to obtain a reliable information on vertebra L-5 (which is an important site of DD), because of the presence of soft tissue disrupting the measurements.

DEXA can only retrieve bone mineral density (BMD), which is not sufficient to determine bone biomechanical properties [6] since BMD does not provide information on bone microstructure nor material properties. Quantitative ultrasound (QUS) techniques represent a complementary approach to DXA [7,8] because ultrasound being mechanical wave, QUS can be used to retrieve bone biomechanical properties.

QUS techniques have been used clinically to estimate bone biomechanical properties at different sites such as the calcaneum [9], phalanx [10], tibia [11], femur [12-14] or radius [15] using transverse [16] and axial [17] transmission devices. However, it remains difficult to assess vertebral bone density due to accessibility issues, the main difficulty consisting in positioning several transducers around the vertebral body during surgery.

Fortunately, when surgery is performed with an anterior approach, metallic cylindrical pins are commonly inserted into the vertebral body in order to maintain soft tissues. The pins are then removed from bone tissue at the end of the surgery. A recent experimental work by our group [18] has suggested to use this pin as a reflector in order to carry out Speed of Sound (SOS) measurements in trabecular bone. It remains difficult to achieve an accurate bone microstructure characterization using usual echographic techniques [8,9,17] because of the strong attenuation and of the complexity of trabecular bone. For instance, multiple scattering phenomena have been evidenced, which makes it difficult to apply classical speckle models to extract quantitative information from the analysis of the backscattered signal. Therefore, the advantage of using a pin as a reflector compared to classical echographic techniques to achieve accurate and reproducible measurements in the echographic configuration. However, the physical determinants of the interaction between a trabecular bone sample including a pin and an ultrasonic wave remain difficult to understand due to the complexity of the phenomena involved. Moreover, the possibility of using normalized broadband ultrasonic attenuation (BUA) to characterize bone quantity has not yet been described in this given configuration. The sensitivity of the QUS parameters to experimental errors related to the relative position of the pin and of the transducer should be determined because it might jeopardize future *in vivo* measurements.

Coupling numerical simulation with high resolution imaging techniques has been shown to constitute a powerful approach capable of bringing further insight on the interaction between bone tissue and ultrasound. In particular, coupling 3-D FDTD simulation tools with 3-D images obtained from microcomputed tomography (μCT) has been employed in the past to model the ultrasonic propagation in trabecular bone [19-21].

The aim of this paper is to develop a numerical model obtained using the coupling of three-dimensional finite difference time domain simulation with X-ray μCT in order to simulate the ultrasonic wave propagation occurring in trabecular bone samples in which a metallic has been inserted. The numerical model aims at taking into account the ultrasonic propagation in trabecular bone as well as the reflection of the ultrasonic wave on the pin (whose axis is perpendicular to the direction of propagation). More specifically, we aim at i) validating the numerical model by comparing the QUS parameters (SOS and BUA) obtained experimentally and numerically, ii) assessing the effect of variation of bone volume fraction on the QUS parameters obtained in this given configuration and iii) assessing the effects of changes of the relative orientation of the pin and of the sensor axis on the QUS parameters.

## II. MATERIALS AND METHOD
### A. Experimental measurements

Twenty-one trabecular bone specimens were obtained from bovine femurs similarly as in reference 18. However, two specimens were excluded from the present study due to a problem which occurred when cutting specimens after the experimental measurements because it was not possible to obtain their 3-D image accurately. As a result, only nineteen specimens were employed in this study, which had already been employed in reference 18. Soft tissues were removed manually using a hand bistoury and the samples were then cut in the proximal region in order to obtain cubic specimens (with a size of approximately 25*25*25mm) using an electric saw.

The ultrasound measurements were carried out similarly as in reference 18. Briefly, all samples were immersed in water at room temperature (which was monitored) and degassed before each measurement. A circular transducer (CMP 89, Sonaxis, Besançon, France) with a resonance frequency equal to 500 kHz and a diameter of the active surface equal to 16.0 mm was positioned in contact with one side of the cubic sample. The ultrasonic transducer was connected to a pulse-receiver amplifier (5052A, Panametrics, Waltham, MA, USA) and used in echographic mode. The received waveform was recorded by a PC via an A/D converter of 100 MHz sampling rate with 12-bit resolution (Spectrum, Grosshansdorf, Germany). The rf signal obtained with the transducer in contact with the bone sample was then recorded and is denoted $S_b(t)$ in what follows.

The same stainless steel pin (Surgiway, Paris, France) as the one used during surgery was then inserted into the bone specimens in a similar manner what is done in the operating room. The pin was a 3.95 mm diameter 240 mm long cylinder made of stainless steel. A clamp was employed to hold the specimens tightly so that no relative movement of the transducer compared to the bone sample is allowed during the insertion of the pin. The distance between the surface of the transducer and the axis of the pin was equal to 11.0 mm. The distance between the pin and the transducer was kept constant thanks to the experimental device shown in Fig. 2 of reference 18, which allows a reproducible insertion of the pin thanks to the presence of a hole with a diameter matching precisely that of the pin. The pin was inserted in this hole, which is rigidly attached to the structure supporting the transducer, when performing the experiments in water as well as when hammering the pin into bone tissue. The rf signal obtained with the pin inserted in the bone sample was then determined and is given by $S_p(t)$ in what follows. The echo of the pin immersed in water was also measured and the rf signal obtained was denoted by $S_w(t)$. As shown in 18, the range of variation obtained for the experimental values BUA, SOS and BV/TV is equal to

18.1 to 43.3 dB/MHz/cm, 1520 to 2200 m/s, and 10.0 to 39.9 %, respectively.

After the ultrasound measurements were performed, the bone samples were cut in the plane perpendicular to the transducer axis and containing the axis of the pin in order to assess bone volume fraction. Only the region of the sample located between the transducer and the pin was considered. The size of the analyzed sample was equal to 10*10*10 mm. An X-ray µCT device (Skyscan1176® scanner, Skyscan, Kontich, Belgium) was used to obtain the 3-D image of the bone specimen with a resolution of 17.7 µm. The CT images were binarized in order to separate the image into bone tissue and liquid. The threshold was chosen so that the value was in the middle of the histogram of CT images. The value of BV/TV (bone volume fraction) was then determined for each sample.

### B. Three-dimensional FDTD numerical simulation

In the framework of the linear elasticity, a 3-D elastic finite-difference time-domain (FDTD) method was used to simulate wave propagation inside the trabecular bone specimens. FDTD methods take into account wave propagation in solids as well as mode conversion. The software packages *SimSonic3D* (available at http://www.simsonic.fr/) was used in this study [22].

Table I shows the material properties used for all media, which were derived by the data of cortical bone and considered as elastic and isotropic [15,23]. A Gaussian pulse with a center frequency equal to 500 kHz and similar to the one used experimentally was employed in the simulation. The parameters shown in Table 1 correspond to material properties at room temperature.

Figure 1 shows an example of the geometrical configuration used in the simulation. The specimen was immersed in water. The original spatial resolution obtained with the X-ray µCT device (17.7 µm) was reduced to 35.4 µm in the simulation model in order to obtain a compromise between a reasonable computation time and an acceptable precision, similarly as what was done in [19,24]. A circular transducer with a diameter equal to 10 mm acting in echographic mode was located in contact with one side of each sample.

The orientation of the specimen and of the direction of propagation was the same as the ones considered experimentally in [18]. In order to avoid wave reflection between the specimen and water at the opposite end of the transducer, the specimen model was duplicated using a planar symmetry relatively to the plane containing the pin axis and perpendicular to the transducer axis, as shown in Fig.1. A cylindrical stainless steel bar with a diameter equal to 4.0 mm was inserted in the bone sample at a similar distance compared to the experimental case.

### C. Determination of the QUS parameters

The QUS parameters were determined following a substitution method described in more details in reference [18]. The same method described below was used for all experimental and simulated rf signals. Examples of wave propagation representation and simulated rf signals are shown in Fig. 2 and 3, respectively.

For each bone sample, SOS and normalized BUA were determined [7] by comparing the *rf* signal obtained from the echo of the pin immersed in water ($S_w(t)$) and the echo of the pin in bone tissue ($S_p(t)$). However, as shown in Fig. 3b-1, it remains difficult to precisely distinguish the wave reflected by the inserted pin from the echoes generated by the microstructure of trabecular bone, which is due to the complex echographic response of the trabecular structure. Therefore, the signal corresponding to the ultrasonic response of bone tissue located between the pin and the transducer is removed by considering the echo of the pin only given by:

$$D(t) = S_p(t) - S_b(t), \qquad (1)$$

where $S_b(t)$ is the echo of the bone tissue without inserting a pin. The determination of $D(t)$ corresponds to a simple way of distinguishing the echographic response of the pin. In what follows, SOS and BUA are determined by comparing $D(t)$ (which corresponds to the echo of the pin immersed in bone tissue) and $S_w(t)$ (which corresponds to the echo of the pin immersed in water, reference signal). SOS was determined following:

$$\text{SOS:} \quad c = \frac{2L}{\frac{2L}{v_{water}} + (t_D - t_w)}, \qquad (2)$$

where $v_{water}$ is the sound speed in water and $L$ is the distance between the surface of the transducer and the pin and $t_D$ and $t_w$ are the time of the first maximum of $D(t)$ and $S_w(t)$, respectively. We chose to consider a time marker in the early part of the signal because such method has been shown to lead to a better correlation between SOS and BV/TV than when using time markers considering the entire signal (such as for example group or phase velocity) [28].

Moreover, normalized BUA was determined by the slope of the curve of the attenuation coefficient $\alpha(f)$ as a function of frequency in the 300-600 kHz range, with

$$\alpha(f) = \frac{1}{2L} 20 \log_{10} \frac{|A_w(f)|}{|A_D(f)|}, \qquad (3)$$

where $A_D(f)$ and $A_w(f)$ are the amplitude spectrum of $D(t)$ and $S_w(t)$, respectively (see Fig. 4).

### D. Effect of a variation of the orientation of the pin.

A potential source of error on the determination of the QUS parameters lies in a possible default of alignment between the transducer axis and the normal of the pin axis. Such effect remains difficult to estimate experimentally because a given bone sample cannot be reused after the pin insertion. Therefore, the effect of a variation of the orientation of the pin relatively to the orientation of the transducer on the determination of the QUS parameters was estimated by modifying the simulation domain. The axis of the pin inserted in trabecular bone was rotated around an axis perpendicular to the ultrasonic propagation for three representative specimens, as shown in Fig. 5. Here, the pin was rotated around an axis in the *y* direction crossing the pin at its center (shown by dots in Fig. 5). The effect of a variation of this tilt angle on the QUS parameters was investigated using virtual models for angles comprised between -16.7 degrees to +16.7 degrees for all samples, which leads to 171 different simulations.

### E. Dilation algorithm

Image processing algorithms were used in order to modify bone volume fraction by using same CT images of a sample. First, the binarizing threshold [29] applied to the original µCT images was varied. In addition to the technique consisting in varying the binarizing threshold, a 3-D dilation algorithm was applied for deriving high BV/TV models (between 50% and 65%) [30-35]: The algorithm replaces the liquid portion, which is surrounded by bone, with bone material so that the bone volume fraction of the model is increased. The use of these image processing algorithms constitutes a simple way of investigating the effect of BV/TV variation, allowing to obtain many different BV/TV values for the same sample. SOS and BUA values were then determined for each modified image.

### III. RESULTS

Figure 6 shows the comparison between experimental and simulated results obtained for SOS and BUA. The correlation coefficient obtained between simulated and experimental SOS was equal to $R = 0.90$ ($p < 0.001$). The correlation coefficient obtained between simulated and experimental BUA is lower but the correlation is significant ($R = 0.55$, $p = 0.015$).

Figure 7 shows the variation of the simulated values of SOS and BUA as a function of BV/TV. A significant linear correlation was obtained between SOS and BV/TV (correlation coefficient $R = 0.82$, $p < 0.001$). However, the linear correlation between BUA and BV/TV was not significant ($p = 0.65$), which will be discussed in section IV.

Figure 8 shows the variation of the experimentally measured values of SOS ($R = 0.77$, $p < 0.001$) and BUA ($p = 0.40$) as a function of BV/TV. Similarly to the simulated results, no significant linear correlation was found between experimental BUA and BV/TV.

Figure 9 shows the variation of the QUS parameters as a function of the incident angle for three representative samples. Figure 9(a) (respectively 9(b)) shows that an error of 6 degrees of the angle causes a variation of around 25~50 m/s for SOS (respectively around 5~10 dB/MHz/cm for BUA).

### IV. DISCUSSION

An originality of this study is to consider the application of a method coupling high-resolution imaging technique and FDTD numerical simulation (which has already been employed in the past with through transmission devices [21,36]) to a configuration corresponding to the reflection of an ultrasonic wave of a metallic pin inserted in trabecular bone. The 3-D FDTD numerical model is validated by comparing experimental and numerical values of SOS. Moreover, we now show the behavior of BUA (both experimental and numerical) as a function of BV/TV and provide physical explanations for the results. Another originality is to assess the sensitivity of the measurement method to variations of the tilt angle, which is important because it provides information on the requirement for precision of the orientation of the pin compared to the transducer axis for future developments of the device. Note that this sensitivity study would not have been possible by considering an experimental approach because the pin insertion can only be done once. More specifically, as shown in Fig. 9, an error of 6 degrees on the tilt angle causes a variation of around 25~50 m/s for SOS (respectively around 5~10 dB/MHz/cm for BUA). Moreover, the experimental errors of SOS (respectively BUA) measurements when realized in a transverse transmission configuration is equal to around 8 m/s (respectively 0.6 dB/cm/MHz) [7]. The comparison of the sensitivity of QUS parameters to the tilt angle and of the reproducibility typically obtained with transverse transmission devices can be used in order to determine the maximum acceptable error on the tilt angle, which is equal to 0.36 - 0.72° for BUA and to 1.0 - 1.8 ° for SOS. The acceptable error on the tilt angle for BUA measurements is significantly lower than SOS measurements, which indicates that the precision requirement is stronger in order to obtain a reliable BUA measurement compared to SOS. This result indicates that SOS may be a more robust indicator than BUA since it may be difficult to practically control the error on the tilt angle with a precision lower than 1°.

### A. Comparison between experiments and simulation

We checked that the range of variation of BUA obtained in the present study (10-50 dB/MHz/cm) is in agreement with previous findings (see for example [37,38]). Different limitations may explain the differences between the experimental and numerical results shown in Fig. 6 concerning both BUA and SOS values. First, bone material properties are not known and generic values are used instead. Second, bone material was assumed to be homogeneous and isotropic. Third, we considered a slightly different geometrical configuration in the simulation compared to the experiments because the diameter of the transducer was smaller (10 mm) in the simulation than in the experiments (16 mm). This choice was made i) in order to take into account the fact that piston mode is not obtained in real transducer and ii) due to a limitation in terms of size of the µCT device. Fourth, the bovine samples used herein have a higher density compared to human vertebral samples (in the range 7%–17%) [39-41]. However, this difference in terms of range of variation in BV/TV does not affect the validity of the results concerning the correlation between BV/TV and the QUS parameters, except that the nonlinear variation of BUA as a function of BV/TV is not likely to be encountered for vertebral bone samples. Following our previous study realized with the same samples [18], we choose not to consider human vertebral samples due to ethical reasons since our aim was to validate the measurement protocol before its application to configurations closer to the clinical situation of interest.

Despite these limitations, a significant correlation was found between numerical and experimental SOS. However, the simulated values of SOS were slightly overestimated compared to the experimental results. The Bland and Altman method [42] revealed a systematic bias (–115.6±50.2 m/s (95% CI)) of the numerically calculated SOS compared to the experimentally measured SOS. The same sources of discrepancies listed above may also hold to explain the difference between numerical and experimental SOS.

The weak correlation obtained between experimental and simulated BUA (which is comparable to the results obtained in reference [43]) can be explained by the limitations described above as well as by the following limitations. In a previous report by Bossy *et al.* [44], a strong linear correlation was evidenced between measured and simulated BUA values with an $R^2$-value of 0.83. One possible explanation for the good correlation found in Bossy *et al.* [44] between experimental and numerical BUA in the case of a classical transverse transmission configuration and the relatively poor correlation found in the present study may be the effect of errors of the tilt angle on the pin compared to the axis of the transducer (see beginning of subsection IV for more details). Errors due to the tilt angle are not present in classical transverse transmission devices and are shown to be more important for BUA than on SOS. Therefore, a possible explanation for the relatively low correlation between experimental and numerical BUA may be the influence of errors on the tilt angle. Another possible explanation for the relatively low correlation between experimental and simulated BUA may be related to phase cancellation effects [45], especially since simulation and experimental transducers had different sizes and therefore different amounts of phase cancellation. Since SOS is computed using a time marker considering the beginning of the signal, SOS may be less sensitive to phase cancellation effects than BUA.

Similarly as in reference [43], our results show that experimental values of BUA were higher than numerical values. Note that the model ignores absorption as a loss mechanism, which may be present in marrow-filled bone specimens [46]. However, as shown in reference [47], the fact that absorption was neglected cannot be used to explain the fact that experimental BUA is higher than theoretical BUA. Note that the effect of absorption on BUA is not understood because contradictory results have been obtained in the literature [47].

### B. Correlation between the QUS parameters and BV/TV

The results shown in Fig. 7(a) indicate a significant correlation between the simulated SOS values and BV/TV, in good agreement with the experimental results obtained in Fig. 4 of the reference [18]. However, a slightly nonlinear relationship is obtained between BV/TV and SOS, with a peak around 30 or 35%, as shown in Fig. 7(a). This result must be interpreted with caution due to the low number of samples considered. Although most studies have found a linear relationship between SOS and BV/TV [e.g.7], some studies have also found nonlinear variations [48], which were explained by the presence of fast and slow waves propagating in the sample. The presence of such phenomena might explain the nonlinear variation of SOS as a function of BV/TV obtained herein. Note that such variations were not observed experimentally nor with the dilation/erosion model because of the particular conditions (in particular in relation with the degree of anisotropy) necessary to distinguish the fast and slow wave modes. Moreover, when fast and slow wave modes can be distinguished, phase cancellation effects [45] may occur, which may be due to a nonlinear variation of the attenuation coefficient as a function of frequency [48] and therefore a modification of the BUA value. Note that in this case, BUA cannot be defined.

Moreover, the variation of BUA as a function of BV/TV obtained experimentally (see Fig. 8b) and numerically (see Fig. 7(b)) is shown to be nonlinear, with a maximum value of BUA for a value of BV/TV of around 25 %. The second order polynomial regression curves are indicated in Fig. 7b and 8b. The maximum value of the regression polynomial fit is comparable for the simulated and the experimental data (31.9 dB/MHz/cm in Fig. 7b and 33.5 dB/MHz/cm in Fig. 8b). Moreover, the maximum value of the regression polynomial fit is obtained for comparable values of BV/TV (25.3% in Fig. 7b and 24.2% in Fig. 8b). Although qualitatively comparable, these results are slightly different than those obtained experimentally by reference [49] who found a maximum value of 60 dB/cm/MHz for a value of BV/TV equal to 18 %. When considering samples with relatively low BV/TV values (between 5 and 10%), increasing BV/TV induces an increase of the size of the trabeculae, which act as scatterers [50,51]. An increase of the size of the trabeculae induces an increase of their scattering cross-section, and thus of the frequency dependent attenuation, which may lead to an increase of the BUA, as obtained in reference [30]. However, a competing effect occurs when BV/TV values are higher (typically around 30 to 40%) since the pores may then act as scatterers instead of the trabeculae for samples with low porosity [52], leading to the opposite behavior of attenuation as a function of porosity. However, more work is needed to understand these phenomena. Note that the modified Biot-Attenborough model and the scattering theory have also been employed in reference [53] to explain the nonlinear variation of BUA as a function of BV/TV.

In order to verify the aforementioned phenomenological explanation, investigations of the effect of BV/TV were performed by modifying *in silico* BV/TV values for three given samples using the erosion/dilation method. To do so, three representative specimens with different BV/TV values were selected (10.0%, 19.4%, and 35.7%, respectively). The results shown in Fig.10 indicate i) a linear increase of SOS as a function of BV/TV for all samples and ii) a strongly nonlinear variation of BUA as a function of BV/TV. The nonlinear variation of BUA as a function of BV/TV (with a maximum value of BV/TV around 20 to 35 %) obtained for all samples constitutes a validation of the phenomenological explanation given above. However, the physiological values of BV/TV obtained for human vertebral trabecular bone is lower than around 17 % [30]. When considering the data shown in Fig.10, Fig.7(b), and Fig.8 for BV/TV values lower than 17%, a significant correlation coefficient is obtained between BUA and BV/TV for simulated ($R = 0.66$) and experimental results ($R = 0.72$). These results are consistent with previous numerical results obtained with human samples [30].

## C. Effect of the tilt angle.

The results shown in Fig. 9 indicate a significant effect of the tilt angle on the QUS parameters for the three representative specimens. Interestingly, the minimum value of SOS and of BUA are obtained for a relatively similar value of the tilt angle for each sample, which is not equal to 0°. These two values of tilt angles, for which minimum values of BUA and SOS are obtained, are noted $TA_{BUA}$ and $TA_{SOS}$ respectively in what follows. Figure 11 shows the relationship between the values of the tilt angles $TA_{BUA}$ and $TA_{SOS}$ for all samples. The values shown in Fig. 11 were obtained using a parabolic regression of the QUS parameters as a function of the tilt angle in order to account for possible numerical errors. One specimen which did not show a concave curve was rejected. A significant correlation ($R = 0.82$) was found between $TA_{BUA}$ and $TA_{SOS}$. Moreover, the linear regression curve is given by the relation $y = 1.16x - 0.41$ and is close to the identity line.

The aforementioned results can be explained as follows. If the ultrasonic propagation occurred in an isotropic medium, the minimum value of SOS would be obtained for a tilt angle equal to zero because it corresponds to a geometrical situation where the propagation distance between the sensor and the pin reaches its minimum value. However, in anisotropic media, the direction of propagation of the ultrasonic energy is not necessarily in the direction perpendicular to the emitter [54]. Let $a$ denote the angle between the normal of the transducer and the direction of propagation of the energy. If $a \neq 0$, the propagation distance will be equal to $D/\cos(aD)$ when the tilt angle is equal to 0 and will be minimum and equal to $D$ when the tilt angle is equal to $a$. Therefore, the anisotropic behavior of bone tissue may explain that the minimum value of the velocity is not obtained for tilt angles equal to 0, as shown in Figs. 9 and 11.

A similar explanation can be used to explain the minimum value of BUA for nonzero tilt angles. If $a \neq 0$, the incident angle of the energy will be equal to $a$ when the tilt angle is equal to 0 and equal to $0$ when the tilt angle is equal to $a$, which explains the higher BUA when the tilt angle is equal to 0.

Therefore, these results may be explained by the structural anisotropy of the samples, which is known to significantly affect the ultrasonic propagation [28]. However, it was not possible to quantify the structural anisotropy of the sample and more work is needed to understand the effect of structural anisotropy on these results. Moreover, the structural anisotropy of trabecular bone affects the received waveforms, which has been evidenced for example in [48] in a through transmission configuration. In some cases, the wave may separate into two waves (fast wave and slow wave, see above), which makes it almost impossible to calculate BUA. In this paper, we selected the geometries where these two wave modes are confounded, which simplifies the analysis of BUA. Moreover, it is impossible to control the direction of wave propagation when performing such measurement in a clinical configuration. For these reasons, we chose not to assess the effect of the structural anisotropy on SOS and BUA.

**D. Future potential *in vivo* measurements.**

The present approach may be used to develop model based inversion approaches to be used in future *in vivo* measurements, in order to eventually lead to possible clinical procedures. A dedicated device adapted to future clinical measurements was developed and is shown in Fig. 12. The device will be inserted around the vertebral body. It is composed of two main bodies (denoted A and B in Fig. 12), which are related by 2 hooks. In clinical practice, the entire device (parts A and B linked by the two closed hooks) will first be inserted between two vertebral bodies after disk resection. Then, the pin will be inserted in the hole located between A and B and will be guided by the device. Then, the ultrasonic measurement will be performed and the hooks will be opened by the surgeon, in order to remove the device, leaving the pin in bone tissue, so that the pin can be used to hold the surrounding soft tissue. Therefore, the pin will not be specifically inserted for the measurement because it will also be used to hold soft tissue, similarly as what is done currently clinical practice. Using the device shown in Fig. 12 allows to insert a pin perpendicular to the transducer axis and positioned at a given distance of the transducer.

The choice of the frequency of 500 kHz was made to obtain a compromise between a "sufficiently high" value of the frequency to obtain a good spatial resolution and a "sufficiently low" value of the frequency in order to obtain acceptable SNR. In particular, when considering the sample with the highest attenuation (50 dB/MHz/cm), assuming a constant frequency dependence of the attenuation coefficient, leads to an attenuation of 110 dB at 1 MHz, which may be too strong compared to usual values of the signal to noise ratio. Therefore, using a value of 1 MHz or higher for the frequency of the transducer could impact the robustness of future *in vivo* measurements, which justifies the choice of 500 kHz for the transducer frequency.

In this study, we used a planar contact transducer, but many studies in the literature employed focused transducers for bone characterization. We chose to consider a planar transducer for the following reasons. First, a planar transducer is adapted to the strong geometrical constraints due to the small space available between the vertebral bodies. Second, using a focused transducer would imply an increased sensitivity on the angle of incidence due to a stronger loss of amplitude when the beam does not intercept the pin axis at its center. Third, we used a contact transducer in order to avoid the reduction of SNR for this preliminary study and to obtain an easy positioning of the transducer relatively to the bone sample, which is important in the measurement configuration.

The presence of cortical bone (which is around 1 mm thick in this region) could disrupt the measurements, due to possible reverberations at the interface between cortical and trabecular bone. This point should be checked in future works. However, the influence of the cortical bone layer could be relatively limited, as shown in 56, which evidenced that the behavior of the received two waves was not modified significantly by the presence of the cortical layer [56]. This paper suggests that the wave propagating in cancellous bone can survive even when it propagates through cortical layer. Moreover, the vertebral body is grinded during surgery. Moreover, the echo of the interface between trabecular and cortical bone arrives much earlier than the echo of the pin. Eventually, the echo of this interface is subtracted, which further decreases the negative impact of this interface.

## V. CONCLUSION

The behavior of the ultrasonic wave reflected by a metallic pin inside cancellous bone was investigated using CT images of cancellous bone specimens. The 3-D FDTD numerical model has been validated by comparing experimental and numerical SOS. Interestingly, the behavior of BUA exhibits non-linear characteristics, with a maximum value for BV/TV values around 20 to 35 %, which corresponds to values higher than what is obtained for human vertebral trabecular bone.

Moreover, the effect of an error on the tilt angle of the pin during the insertion procedure was investigated. The results showed that the angle of the pin strongly affects the value of BUA while it weakly affects the value of SOS. These results indicate that the proposed method using the ultrasonic wave reflected by a pin has a potential to estimate the bone density although the data is sensitive to the geometrical properties of the configuration obtained during the insertion protocol. Moreover, SOS seems to constitute a more reliable parameter compared to BUA due to its lower sensitivity of errors on the tilt angle and to its linear variation with BV/TV.

Table

Table I: Material properties used for bone tissue (generic values taken from [e.g. 7]) and for stainless steel (from [e.g. 55]) in the simulations.

| Material | | Wave Speed [m/s] | Density [kg/m$^3$] |
|---|---|---|---|
| Bone | Longitudinal | 4,000 | 2,000 |
| | Shear | 2,000 | |
| Water | | 1,483 | 1,000 |
| Stainless steel | Longitudinal | 5,882 | 7,900 |
| | Shear | 3,144 | |

Figures

Figure 1: An example of the geometrical configuration of three-dimensional simulation. Figures (a)-(c) show cross-section diagrams and (d) shows the three-dimensional view. (Color online)

Figure 2: Examples of the screenshots of the simulated wave distribution. Figure (a-1) to (a-3) show the results with pin and (b-1) to (b-3) shows without pin. (Color online)

Figure 3: Examples of the simulated rf signals (a) without bone specimen and (b) with bone specimen. (b-1): original rf signals and (b-2): difference of the signals shown in (b-1) and namely the signals obtained with the pin and without the pin. (Color online)Figure 4: (a): normalized spectrum of the simulated rf signals and (b): frequency dependence of the attenuation coefficient derived using the two spectra shown in (a). (Color online)

Figure 5: Simulation setup used for the evaluation of the effect of a variation of the pin orientation. The pin axis was rotated around an axis perpendicular to the ultrasonic

propagation. (Color online)

Figure 6: Comparison between experimental and simulated results obtained for SOS and BUA.

Figure 7: Variation of the simulated values of SOS and BUA as a function of BV/TV.

Figure 8: Variation of the experimentally measured values of SOS and BUA as a function of BV/TV.

Figure 9: Variation of the QUS parameters as a function of the incident angle for three different samples. (Color online)

Figure 10: Variation of the QUS parameters as a function of *in silico* BV/TV values when a dilation/erosion process was applied to three specimens in order to modify bone volume fraction by adding/suppressing bone voxel at the surface of trabecular bone. (Color online)

Figure 11: Relationship between the values of the tilt angle for which the values of BUA and SOS are minimum.

Figure 12: A dedicated device adapted to future clinical measurements. (Color online)

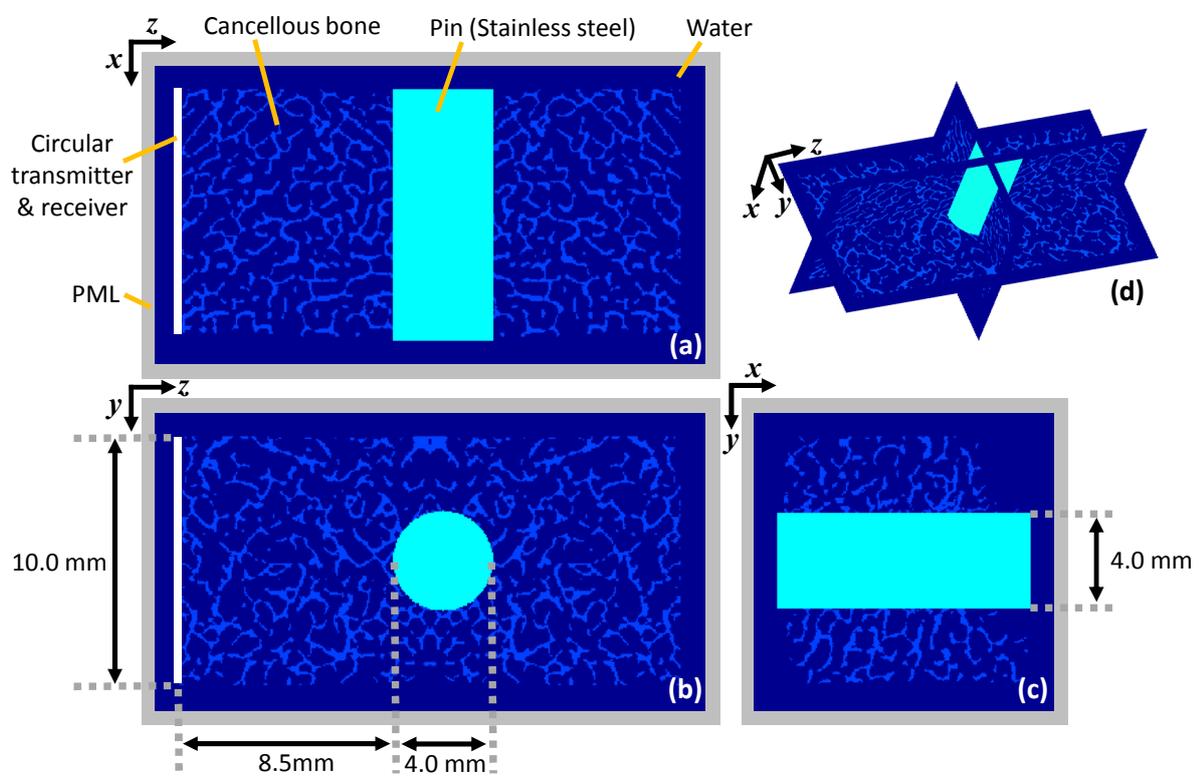

Fig.1

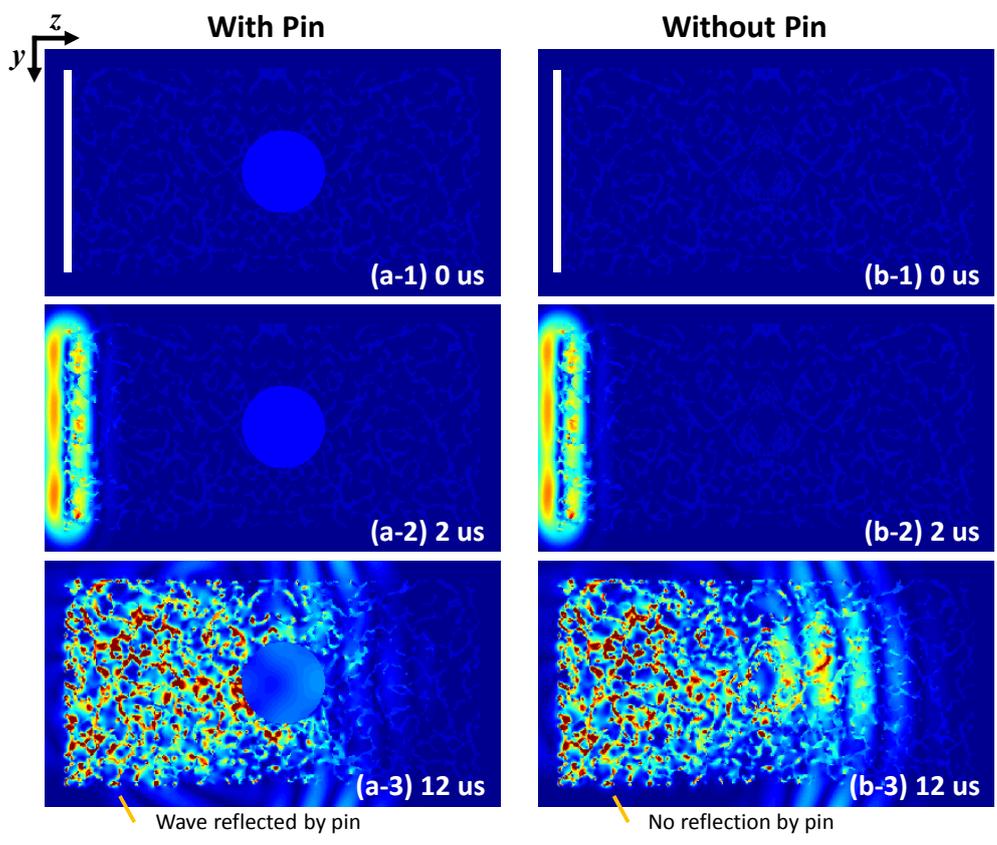

**Fig.2**

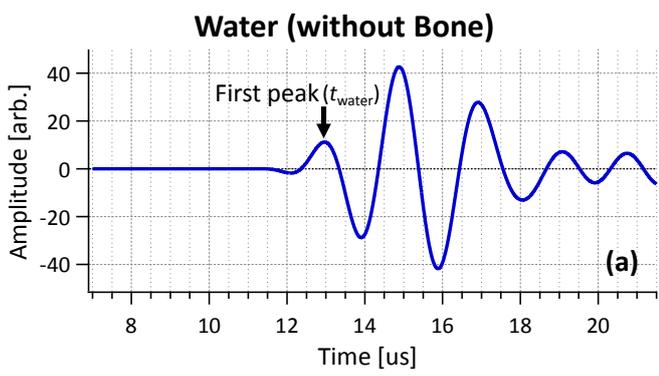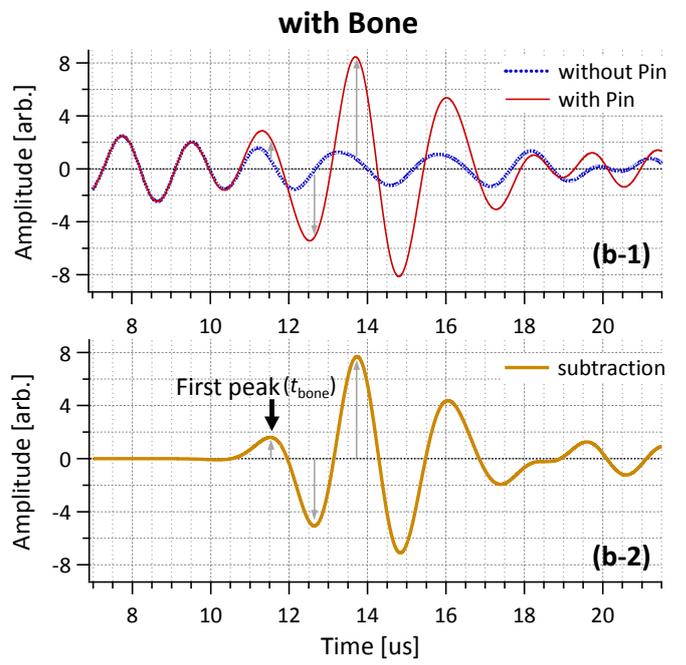

Fig.3

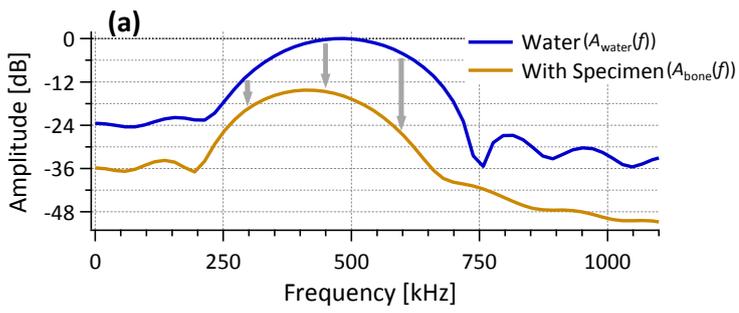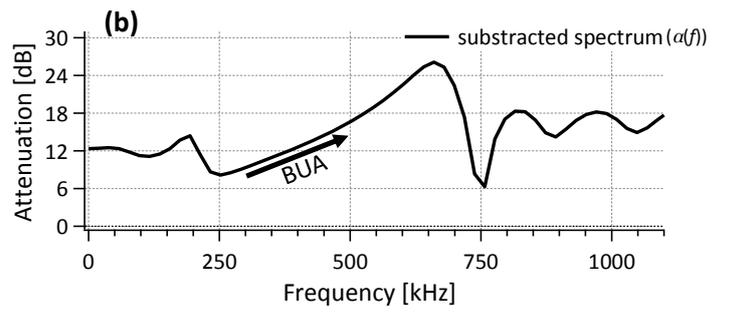

**Fig.4**

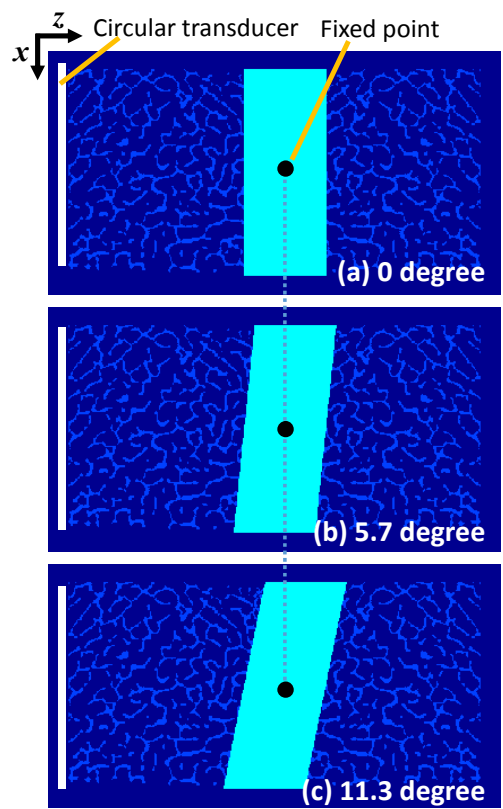

Fig.5

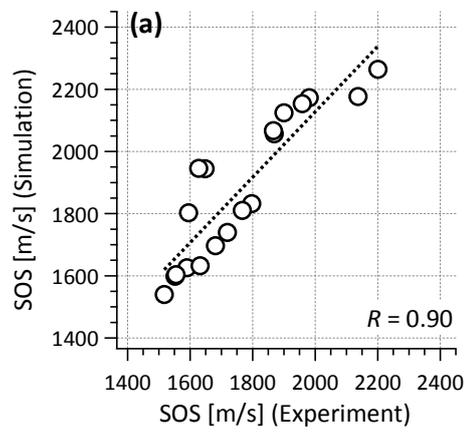 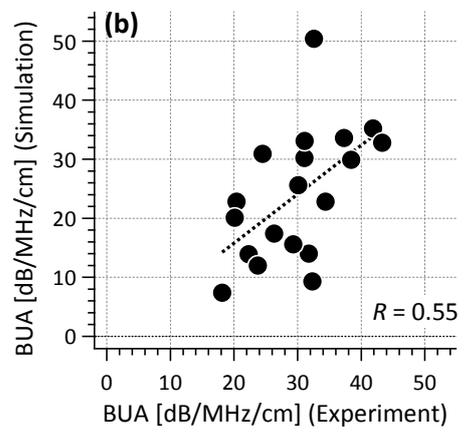

**Fig.6**

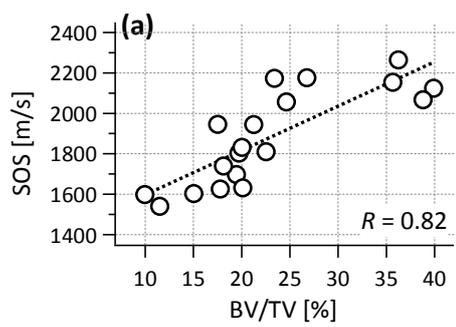 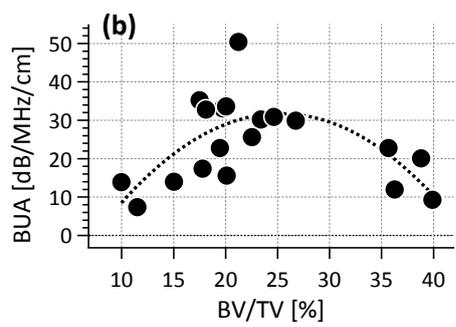

**Fig.7**

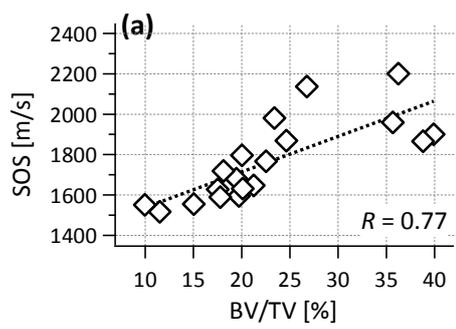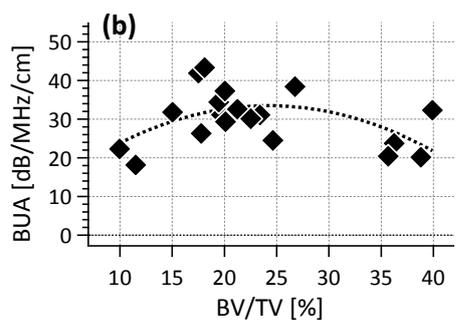

**Fig.8**

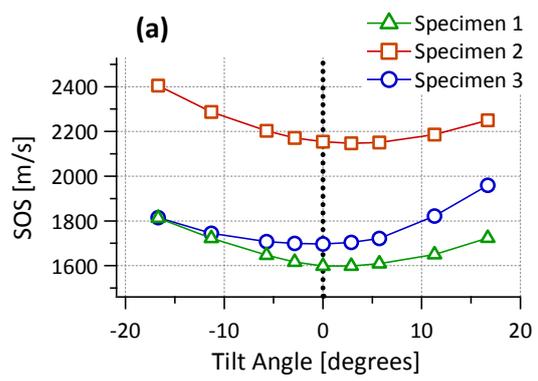 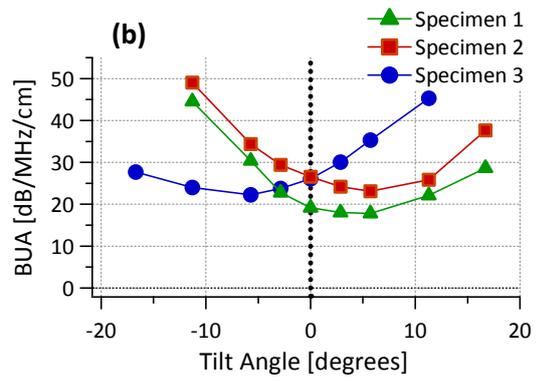

**Fig.9**

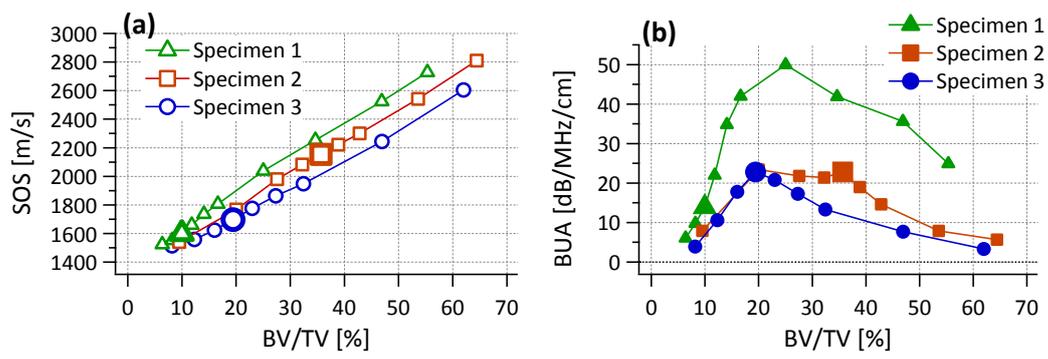

**Fig.10**

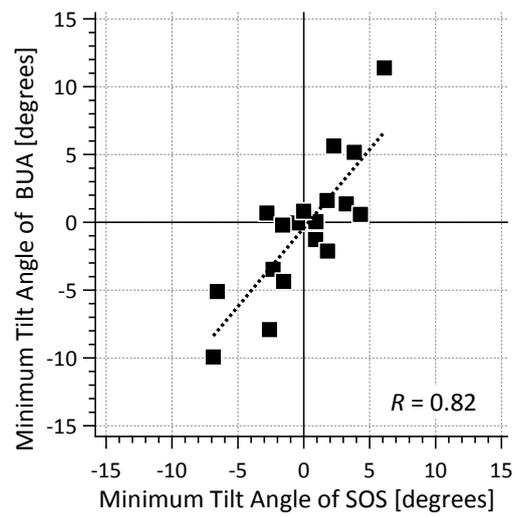

**Fig.11**

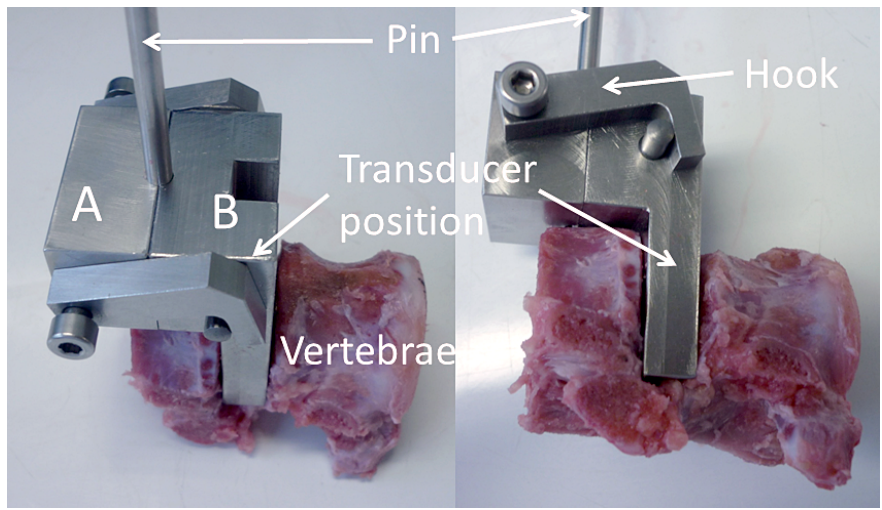

**Fig.12**

| Material | | Wave Speed [m/s] | Density [kg/m³] |
|---|---|---|---|
| Bone | Longitudinal | 4,000 | 2,000 |
| | Shear | 2,000 | |
| Water | | 1,483 | 1,000 |
| Stainless steel | Longitudinal | 5,882 | 7,900 |
| | Shear | 3,144 | |

**Table I**